# Positive unidirectional anisotropy in $Y_3Fe_5O_{12}/Ir_{20}Mn_{80}$ bilayers

E. C. Souza[1*], P. R. T. Ribeiro[2], J. E. Abrão[1], F. L. A. Machado[1], and S. M. Rezende[1]

[1]Departamento de Física, Universidade Federal de Pernambuco, 50670-901, Recife, Pernambuco, Brazil.

[2]Instituto de Física Gleb Wataghin, Universidade Estadual de Campinas, 13083-859, Campinas, São Paulo, Brazil.

ABSTRACT

We report an experimental study of the unidirectional anisotropy in bilayers made of the important ferrimagnetic insulator yttrium iron garnet (YIG) and the room temperature antiferromagnet $Ir_{20}Mn_{80}$ (IrMn). Measurements of the magnetization hysteresis loop in a wide temperature range and ferromagnetic resonance at room temperature revealed an unconventional positive exchange bias (EB). For comparison, we also made FMR measurements in a Py/IrMn bilayer that led to a negative EB with amplitude nearly two orders of magnitude larger than in YIG/IrMn. The presence of the positive EB, in which the hysteresis loop shift occurs in the direction of the field applied during deposition of the films, is attributed to an antiferromagnetic coupling between the spins of the two layers at the interface. The small value of the EB field in YIG/IrMn can be attributed to a competition of the interactions between the spins of the two sublattices of antiferromagnetic IrMn and ferrimagnetic YIG produced by the interface roughness.

*Corresponding author: E-mail: edycleyson.souza@ufpe.br

# I. INTRODUCTION

The phenomenon of exchange bias (EB), also called exchange anisotropy, or unidirectional anisotropy, results from the interfacial magnetic interactions at the atomic contacts between ferromagnetic (FM) and antiferromagnetic (AF) materials. Discovered over six decades ago [1], this phenomenon has been widely investigated theoretically and experimentally by means of several techniques and has also found applications in magnetoelectronic and sensor devices [2-6]. In FM/AF thin film bilayers the most known manifestation of the EB is the shift of the hysteresis loop along the field axis by some value $H_{eb}$ that varies inversely with the thickness of the FM layer. Usually, the exchange bias direction is opposite to the direction of the magnetic field applied during the sample growth, and is called negative EB. The reverse situation in which the hysteresis loop shifts in the direction of the field applied in the sample growth, called positive EB, was first observed in bilayers of $Fe/FeF_2$ [7] and is usually associated with an antiferromagnetic interfacial coupling at the FM/AF interface [6,8,9]. In most systems the size and sign of the exchange bias field can be engineered with appropriate thermal-field treatments in the so-called training effect, that consists in performing several consecutive hysteresis loops with adequate temperatures and applied field amplitudes [6-9].

One material that has played a key role in magnetism is the ferrimagnetic insulator yttrium iron garnet, known as YIG, that has chemical formula $Y_3Fe_5O_{12}$. Due mainly to its very low magnetic and acoustic losses and long spin-lattice relaxation time, for several decades YIG has been the prototype material for studies of the basic physics of a variety of magnonic phenomena [10-16]. In the last decade YIG has gained tremendous attention as a key material in insulator-based magnon spintronics for revealing unique features of various effects, such as the spin-pumping and spin-Hall effects [17-21], spin-Seebeck, spin-Peltier and other thermal effects [22-28], as well as spin-Hall magnetoresistance [29,30]. Despite its importance for magnetism, there are very few studies of the exchange bias phenomenon involving YIG [31,32]. One of the reasons for this is that the exchange bias field decrease rapidly with increasing FM film thickness, and most studies with YIG have been made with bulk samples cut from crystals grown by the floating zone method and flux growth [33], or relatively thick films prepared by liquid phase epitaxy [34]. Only in recent years with the advent of techniques to fabricate nano-thick YIG films [35] the study of EB phenomena with YIG/AF became feasible.

In this paper we present an experimental study of the unidirectional anisotropy in FM/AF bilayers made of YIG and $Ir_{20}Mn_{80}$ (IrMn) employing the techniques of magnetization hysteresis loop measurements and X- band microwave ferromagnetic resonance (FMR). The measurements in a wide range of temperature revealed an unconventional positive exchange bias (EB). For comparison, we also made FMR measurements in a Py/IrMn bilayer that led to a negative EB with amplitude nearly two orders of magnitude larger than in YIG/IrMn. The small value of the EB field is attributed to a competition of the interactions between the spins of the two sublattices of antiferromagnetic IrMn and ferrimagnetic YIG produced by roughness at the FM/AF interface.

## II. SAMPLE FABRICATION AND CHARACTERIZATION

The bilayers of $Y_3Fe_5O_{12}$(30-100 nm)/$Ir_{20}Mn_{80}$(74 nm) used in this investigation were prepared by RF and DC magnetron sputtering techniques. The YIG layers were deposited with the DC technique on commercial (111)-oriented $Gd_3Ga_5O_{12}$ (GGG) substrates with lateral dimensions 1.5 ✕ 3.0 mm² and thickness 0.5 mm. Prior to deposition, the GGG substrates were cleaned in an ultrasonic bath of acetone and isopropyl alcohol for 10 min. The base pressure was 3 ✕ $10^{-7}$ Torr and the deposition pressure was of 3.0 mTorr with 550 sccm argon flow inside the sputtering chamber, resulting in a deposition rate of 2.52 nm/min. After deposition, the YIG films were annealed in $O_2$ atmosphere at 800 °C for 4 hours with heating rate of 10 °C/min and then cooled with cooling rate 1 °C/min. The AF layer of $Ir_{20}Mn_{80}$ (denoted by IrMn) was deposited by DC sputtering over the YIG film magnetized by a magnetic field of 340 Oe applied in the sample plane. The base pressure was 3 ✕ $10^{-7}$ Torr and the deposition pressure was of 3.0 mTorr with 550 sccm argon flow, providing a deposition rate of 7.4 nm/min. The bilayer of $Ni_{81}Fe_{19}$ (Permalloy- Py) and IrMn used for reference was prepared by DC sputter in a similar way.

The structural properties were measured through x-ray diffraction (XRD) using a Rigaku x-ray diffractometer, model Smartlab, with the $CuK_\alpha$ radiation (λ = 1.5418 Å) in a $\theta$-$2\theta$ Bragg-Brentano geometry. Figure 1 (a) shows a sketch of the $Y_3Fe_5O_{12}(t)$/$Ir_{20}Mn_{80}$(74 nm) bilayers with $t$ = 30, 50, 80, and 100 nm. The interface zoom of FM/AF layers illustrates the spin arrangement in the antiferromagnetic coupled interface, a feature of positive exchange bias [33]. Figure 2 (b) shows an XRD spectrum for the $Y_3Fe_5O_{12}$(100 nm)/$Ir_{20}Mn_{80}$(74 nm) sample. For the YIG layer we observe the presence of peaks at $2\theta \approx 25°$ and $2\theta \approx 51°$ associated with the YIG (222) and YIG (444) planes, suggesting an epitaxial growth in the GGG substrate attributed to $K_{\alpha 1}$ and $K_{\alpha 2}$ lines of the incident x-ray. Beside the YIG peaks, the peak observed at $2\theta \approx 41.5°$ is due to the IrMn (111) FCC texture. Previous studies have shown that this texture is a striking feature of a good exchange bias effect [34-36]. No other distinct peaks are not observed.

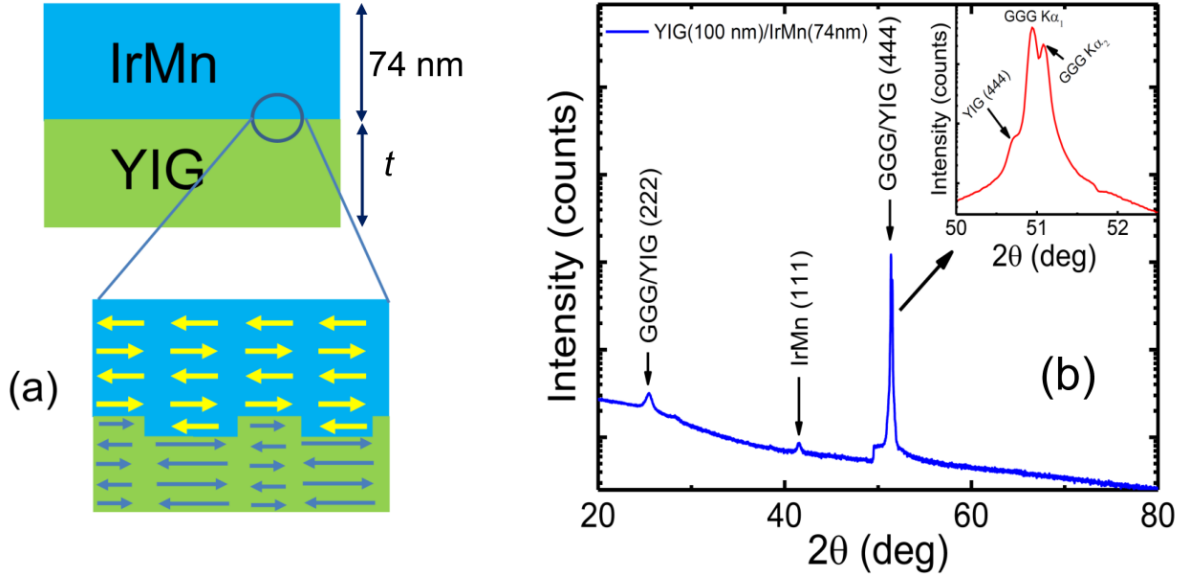

**Figure 1** (Color online) (a) Sketch of the bilayer sample structure and spin arrangement of antiferromagnetically coupled YIG and IrMn layers. (b) XRD $\theta$-$2\theta$ scan of a IrMn(74 nm)/YIG(100 nm)/GGG sample showing the peaks corresponding to each layer and the GGG substrate.

### III. MAGNETIZATION MEASUREMENTS

The hysteresis loops were measured using a superconducting quantum interference device (SQUID) based magnetometer (Quantum Design model MPMS-3) vibrating-sample magnetometer, with precision $< 1 \times 10^{-8}$ emu ($\leq 2500$ Oe) and $< 8 \times 10^{-8}$ emu ($> 2500$ Oe). In order to avoid contributions due to residual fields, the magnet was demagnetized in an oscillation mode before each measurement. In addition, a paramagnetic standard Pd sample was used as a reference material for calibration with residual field ∼1 Oe.

The magnetic field $H$, applied on the sample plane along the direction of the field used to magnetize the YIG film during the deposition of the IrMn layer, was swept between ±30 Oe. The sweep procedure starts from +30 Oe and descending intensity to −30 Oe and returns to +30 Oe with increments of 1 Oe, keeping the temperature fixed. In all hysteresis loops the paramagnetic contribution from GGG was subtracted. Figure 2 shows the hysteresis loops measured at room temperature in all YIG (*t*)/IrMn samples with thicknesses *t* = 30, 50, 80, and 100 nm. The exchange bias field $H_{eb}$ is maximum in the 30 nm sample, shown in Fig. 2(a), and decreases with increasing YIG thickness, as expected, and becomes negligible for the sample with *t* = 100 nm (Figure 2(d)). The values of $H_{eb}$ measured by the hysteresis loop shift (HLS) in the four samples are listed in Table 1. The inset in Figure 2(d) shows the typical decay of $H_{eb}$ with increasing YIG film thickness of the FM layer [3].

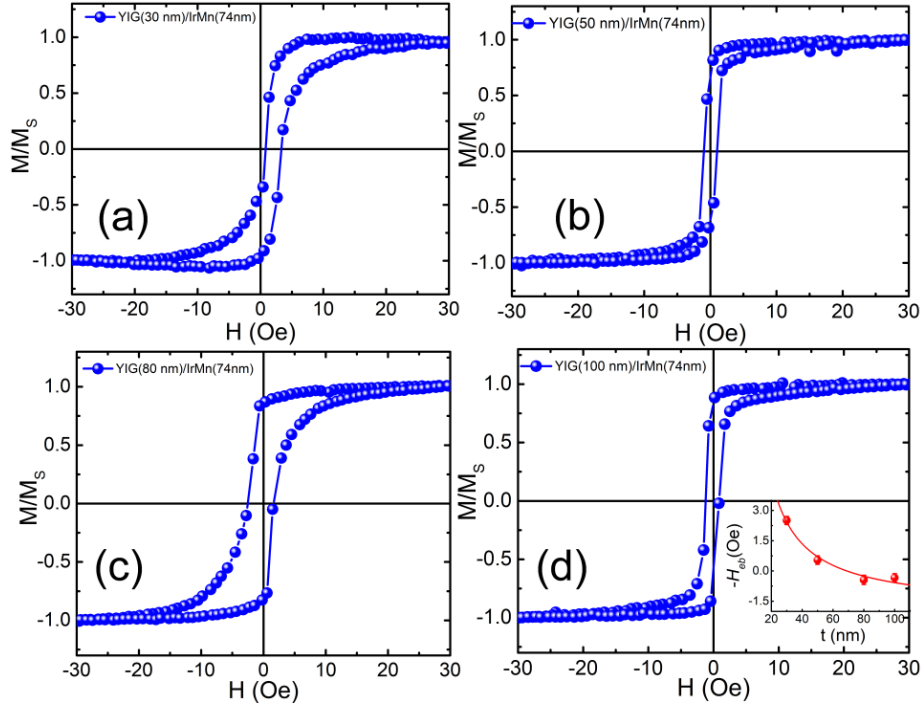

**Figure 2** (Color online) Magnetization loops for YIG(*t*)/ IrMn(74 nm), *t* = (a) 30 nm, (b) 50 nm, (c) 80 nm, and (d) 100 nm. All measurements were done at room temperature.

The magnetization loops of the YIG (30 nm)/IrMn(74 nm) sample for several temperatures are shown in Fig. 3(a). At all temperatures one clearly sees a field shift in the hysteresis loop and enhanced coercivity, characteristic of exchange bias phenomena. The physical origin of the shift resides in the fact that the AF spin arrangement breaks up into domains, having a net moment along the direction of the cooling field, that couples antiferromagnetically with the spins of the two sublattices of ferrimagnetic YIG. Figure 3(b) shows the temperature dependencies of the coercive field $H_c$ and the exchange bias field $H_{eb}$, exhibiting the typical decay of both fields with increasing *T*.

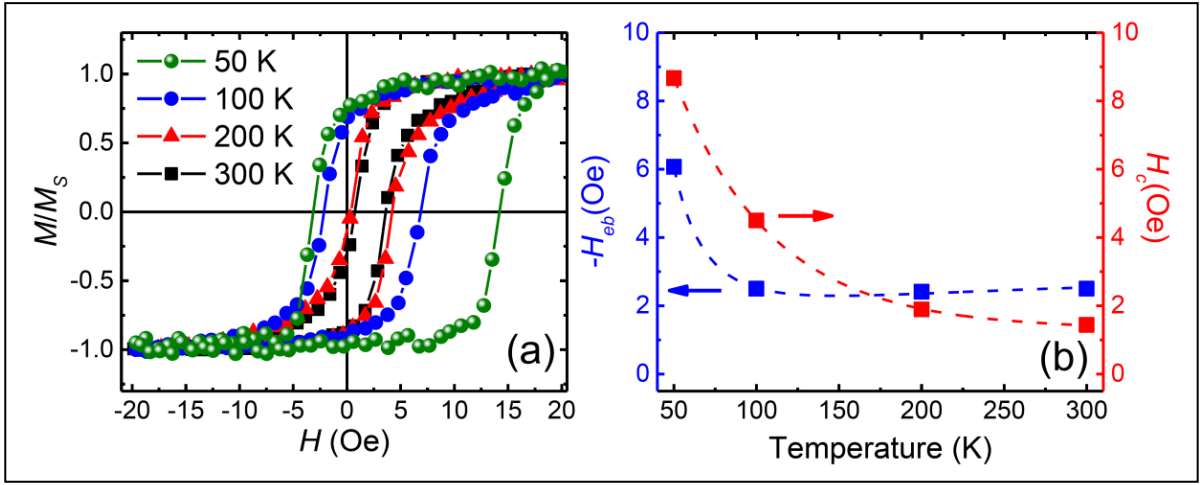

**Figure 3** (Color online) (a) Magnetization loops of YIG (30 nm)/IrMn(74 nm) for $T$ = 50, 100, 200, 300 K. (b) Exchange bias field $H_{eb}$ and coercive field $H_c$ as a function of temperature.

## IV. FERROMAGNETIC RESONANCE MEASUREMENTS

Ferromagnetic resonance measurements were carried out in a homemade X-band microwave spectrometer. The sample is mounted on the tip of PVC rod and inserted into a rectangular microwave cavity operating in the $TE_{102}$ mode, with resonance frequency of 9.417 GHz, quality factor of 2000 and an incident rf power $P_{rf}$ = 20 mW, low enough to avoid nonlinear excitation of spin waves. In order to avoid rf electric effects, the samples were placed close to the end wall of the cavity, in a nodal position of minimum rf electric field and maximum rf magnetic field perpendicular to the static field, such that the rf magnetic field drives the magnetization precession. This precaution avoids the generation of galvanic effects caused by the rf electric field. With this arrangement the bilayer sample can be rotated in the plane so as to measure the angular dependence of the resonance field. The static magnetic field produced by the poles of an electromagnet was modulated at 1.2 kHz with a pair of Helmholtz coils to allow lock-in detection of the absorption derivative.

Figure 4 shows FMR absorption curves measured in YIG($t$) and YIG($t$)/IrMn(74 nm) samples with the static field applied in the direction $\phi_H$ = 0°, where $\phi_H$ = 0 is the in-plane direction of the field applied during the deposition of the AF layer. The red solid lines correspond to numerical fits of the experimental data with derivatives of a Lorentzian function, from which we obtain the FMR field $H_R$ and the linewidth $\Delta H$. All YIG films show linewidths that decrease with increasing film thickness, varying from 15.5 Oe to 9.3 Oe, due to two-magnon scattering [15]. After deposition of the IrMn layer, all samples show an additional increase in linewidth, which is a signature of the spin pumping effect [19]. Also, after deposition of the IrMn layer onto YIG, we observe an upward shift in the FMR field for all samples, which is attributed to a negative rotatable anisotropy [36-42].

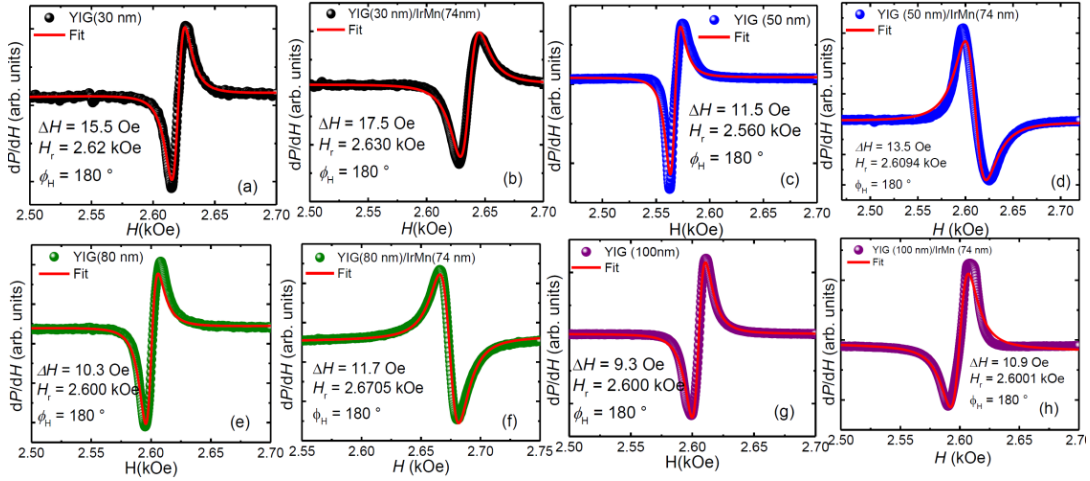

**Figure 4** (Color online) FMR absorption curves for the YIG (*t*) samples measured at 9.417 GHz before and after deposition of the IrMn layer. The solid lines represent the numerical fits with Lorenztian derivative functions that give the values of the linewidth Δ*H* and field for resonance $H_R$.

Figures 5(a-d) show the in-plane angular dependencies of the FMR fields measured in the $Y_3Fe_5O_{12}(t)/Ir_{20}Mn_{80}(74$ nm) bilayers with *t* = 30, 50, 80, and 100 nm with microwave frequency of 9.417 GHz. The red solid lines correspond to the best fits to data using the theory to be described, from which the values of the exchange bias field, effective saturation magnetization, uniaxial and rotational anisotropy fields were extracted. For comparison, we show in Figs. 5(e,f) the measurements made in a 30 nm thick film of permalloy (Py- $Ni_{81}Fe_{19}$) and in a Py(30 nm)/$Ir_{20}Mn_{80}$(74 nm) bilayer.

In order to interpret the experimental data, we need an expression for the FMR frequency of a FM film with exchange bias. For this, consider the energy per unit volume for the FM magnetization $\vec{M}$ [36]

$$E = -\vec{H}\cdot\vec{M} - \frac{K_u}{M^2}\left(\hat{u}\cdot\vec{M}\right)^2 + 2\pi\left(\hat{n}\cdot\vec{M}\right)^2 - \frac{K_S}{M^2 t}\left(\hat{n}\cdot\vec{M}\right)^2 - \vec{H}_{eb}\cdot\vec{M} - \vec{H}_{RA}\cdot\vec{M}, \quad (1)$$

where the various terms represent, in order from the left, Zeeman, uniaxial anisotropy, demagnetizing, surface anisotropy, exchange bias, and rotational anisotropy [39]. The symbols in Eq. (1) are: $K_u$ is the parameter for the uniaxial anisotropy in the direction $\hat{u}$ of the film; $K_S$ is the parameter for the surface anisotropy of the film with thickness *t*; $\vec{H}_{eb}$ is the exchange bias field, and $\vec{H}_{RA}$ is the rotational anisotropy field [36]. Consider that the film is in a *x-y* plane of a coordinate system with the external static field *H* in the film plane at an angle $\phi_H$ with the *x* axis, and that the exchange bias field is along $\phi_H = 0$. The energy in Eq. (1) can be written in terms of the angle coordinates from which the FMR frequency can be calculated with the following expression [43]

$$\omega^2 = \frac{\gamma^2}{M \sin^2 \theta_0} \left[ \frac{\partial^2 E}{\partial \theta^2} \frac{\partial^2 E}{\partial \phi^2} - \left( \frac{\partial^2 E}{\partial \theta \partial \phi} \right)^2 \right]_{\theta_0, \phi_0}, \qquad (2)$$

where $\theta_0$ and $\phi_0$ are the angles of the equilibrium magnetization $\vec{M}$, given by $\partial E / \partial \theta = \partial E / \partial \phi = 0$. For $\theta_H = \pi / 2$ one finds $\theta_0 = \pi / 2$, while the azimuthal angle $\phi_0 \approx \phi_H$ for applied fields much larger than the exchange bias and anisotropy fields.

Using in Eq. (2) the second derivatives of the energy in Eq. (1) expressed in terms of the angular coordinates, we obtain for the FMR frequency of the film with exchange bias

$$\omega_0 = \gamma [H + H_u \cos^2 \phi_H + H_{RA} + H_{eb} \cos \phi_H)]^{1/2} [H + H_u \cos^2 \phi_H + H_{RA} + H_{eb} \cos \phi_H + 4\pi M_{eff}]^{1/2},$$

(3)

where $\gamma$ *is* the gyromagnetic ratio, $H_u = 2K_u/M$ is the uniaxial anisotropy field considered to point in the direction $\phi = 0$, and $4\pi M_{eff} = 4\pi M - H_S$ is the effective magnetization, where $H_S = 2K_S/tM$ is the surface anisotropy field. Solution of this equation gives the field for resonance at a frequency $\omega_0$

$$H_R = -2\pi M_{eff} + \left[ (2\pi M_{eff})^2 + (\omega_0 / \gamma)^2 \right]^{1/2} - H_u \cos^2 \phi_H - H_{RA} - H_{eb} \cos \phi_H. \qquad (4)$$

The solid lines in Fig. 5 represent fits to the FMR data with Eq. (4) from which we obtain the values for the exchange bias, uniaxial anisotropy, and rotational anisotropy fields, as well as the effective magnetization, as presented in Table 1. The value of the exchange bias field decreases with increasing thickness of the YIG film, as expected. Also, the value of the effective magnetization increases with increasing thickness and approaches the value of the bulk saturation magnetization, $4\pi M_{eff} = 1.77$ kG, because the perpendicular anisotropy varies with $1/t$. Notice also that the values for the exchange bias field measured by FMR are different from the field shifts of the hysteresis curve. This is attributed to the sensitivity of the exchange bias phenomenon to the measurement technique, an intriguing and still not completely resolved question [36,44].

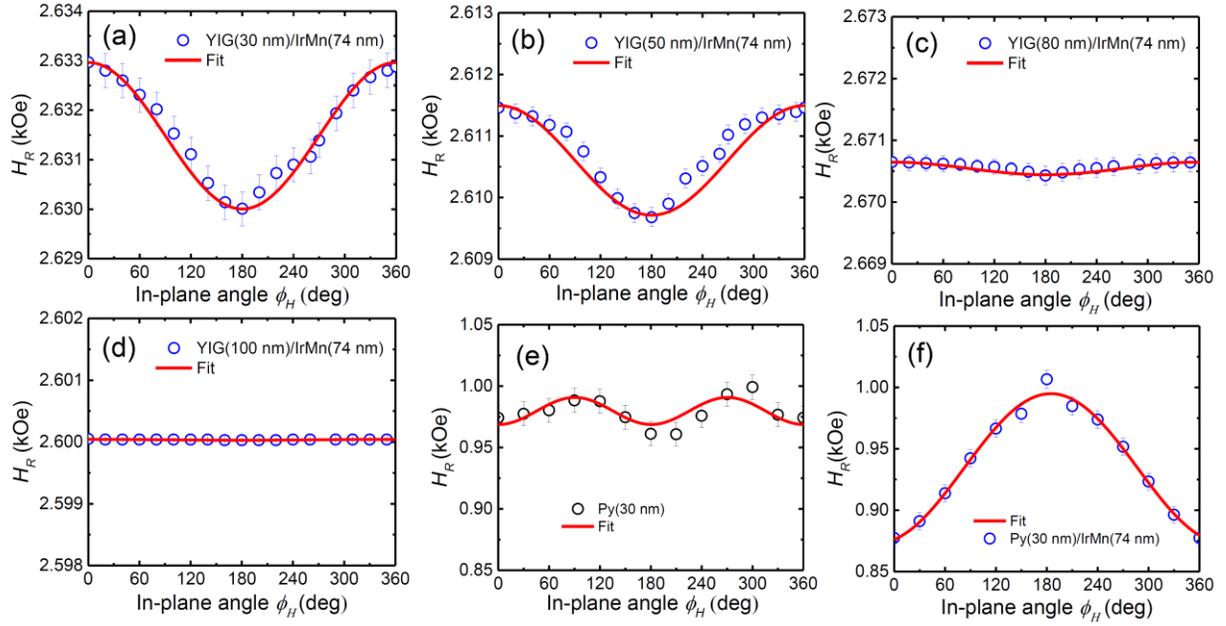

**Figure 5** (Color online) In-plane angular dependence of the resonance field $H_R$ measured at 9.417 GHz in the following samples. (a)-(d) YIG ($t$)/IrMn (74 nm) bilayers with $t = 30, 50, 80,$ and 100 nm; (e) Py (30 nm); and (f) Py (30 nm)/IrMn (74 nm) bilayer. The solid lines represent the numerical fits with Eq. (4) that give the values of the parameters in Table 1.

| Bilayer | YIG(30 nm)/IrMn | YIG(50 nm)/IrMn | YIG(80 nm)/IrMn | YIG(100 nm)/IrMn | Py(30 nm)/IrMn |
|---|---|---|---|---|---|
| $H_{eb}$ (Oe)-HLS | -2.5 | -0.54 | 0.4 | 0.3 | -- |
| $H_{eb}$ (Oe)-FMR | -1.5 | -0.89 | -0.22 | -0.01 | 65.0 |
| $4\pi M_{eff}$ (kG) | 1.65 | 1.71 | 1.71 | 1.77 | 11.0 |
| $H_{RA}$ (Oe) | 7.2 | 6.9 | 6.8 | -0.61 | 2.0 |
| $H_u$ (Oe) | 0 | 0 | 0 | 0 | 11.0 |

**Table 1.** Parameters obtained from the fits of Eq. (4) to the FMR data for YIG/IrMn and Py/IrMn bilayer samples.

For the purpose of comparing the data in YIG/IrMn with a well studied system, we have made FMR measurements in a permalloy film with the same thickness as the thinnest YIG film, with and without a IrMn (74 nm) layer. The angular dependencies of the FMR field of the bare Py film and the Py/IrMn bilayer are shown in Figs. 5(e,f). Clearly, while YIG/IrMn has a positive exchange bias, in Py/IrMn the exchange bias is negative, which is the most common case in FM/AF bilayers. The fit of Eq. (4) to the data in Fig. 5(f) gives an exchange bias field of 65.0 Oe, which is about 40 times larger than the value measured in the bilayer with the YIG film of the same thickness.

## V- CONCLUSIONS

In summary, we have investigated the unidirectional anisotropy in YIG/IrMn bilayers by means of magnetization hysteresis loop and ferromagnetic resonance measurements. Both techniques revealed an unconventional and relatively modest positive exchange bias (EB) in this system. For comparison we also made FMR measurements in a Py/IrMn bilayer, a well studied system, which led to a negative EB with amplitude nearly two orders of magnitude larger than in YIG/IrMn. The presence of the positive EB, in which the hysteresis loop shift occurs in the direction of the field applied during deposition of the films, is attributed to an antiferromagnetic coupling between the spins of the two layers at the interface. The small value of the EB field in YIG/IrMn can be attributed to a competition of the interactions between the spins of the two sublattices of antiferromagnetic IrMn and ferrimagnetic YIG produced by the interface roughness, as illustrated in Fig. 1(a). While the AF coupling in one area produces an effective field on the spins of YIG that points in a certain direction, in another area the effective field points in the opposite direction, resulting in a reduced exchange bias.

**ACKNOWLEDGEMENTS**

This work was supported by the Brazilian agencies Conselho Nacional de Desenvolvimento Científico e Tecnológico (CNPq), Coordenação de Aperfeiçoamento de Pessoal de Nível Superior (CAPES), Financiadora de Estudos e Projetos (FINEP), Fundação de Amparo à Pesquisa do Estado de São Paulo (FAPESP) Grant No. 2022/04496-0 and Fundação de Amparo à Ciência e Tecnologia do Estado de Pernambuco (FACEPE).